\documentclass[pdf-a,balance,colorlinks,upint,subscriptcorrection,varvw,mathalfa=cal=boondoxo, spanish,french,vietnamese,russian,greek]{asmeconf}

\hypersetup{%
	pdfauthor={John H. Lienhard},									  % <=== change to YOUR name
	pdftitle={ASME Conference Paper LaTeX Template},                  % <=== change to YOUR pdf file title
	pdfkeywords={ASME conference paper, LaTeX template, BibTeX style},% <=== change to YOUR pdf keywords
	pdfsubject = {Describes the asmeconf LaTeX template},			  % <=== change to YOUR subject
	pdflicenseurl={https://ctan.org/pkg/asmeconf},% may delete
}

\usepackage{algorithm} 
\usepackage{algpseudocode}

\begin{document}

% Change these fields to the right content for your conference.
% You can comment these out if for some reason you don't want a header.
% Use title case (first letters capitalized), not all capitals

\ConfName{Proceedings of the ASME 2024\linebreak International Design Engineering Technical Conferences and\linebreak 
Computers and Information in Engineering Conference}
\ConfAcronym{IDETC/CIE2024}
\ConfDate{August 25--28, 2024} % update 
\ConfCity{Washington, DC} % update 
\PaperNo{DETC2024-143139}

% Units of measure (e.g., cm) and other specialty lowercase terms in the title should be 
%   enclosed in \NoCaseChange{...} to maintain lower case type
%   LaTeX will automatically set the rest of the title in all capital letters.

\title{Prototype2Code: End-to-end Front-end Code Generation from UI Design Prototypes} % <=== replace with YOUR title
%\title{Place Title Here: Place Subtitle After Colon} 
 
%   Put author names into the order you want. Use the same order for affiliations.
%   \affil{#} tags the author's affiliation to the address in \SetAffiliation{#}.
%   No space between last name and \affil{#}, separate names with commas.
%
%	For a sole author or a single affiliation for all authors, {#} may be left empty, as \affil{} and \SetAffiliation{} (but not with [grid] option!)
%
%   \CorrespondingAuthor{email} follows that author's affiliation, no spaces.  
%   If multiple corresponding authors, put both email addresses in the same command and place after both authors.
%
%   \JointFirstAuthor, if applicable, follows the affiliation of the relevant authors, no spaces.

\SetAuthors{%
        Shuhong Xiao\affil{1},
        Yunnong Chen\affil{1},
        Jiazhi Li\affil{1},
        Liuqing Chen\affil{1}\affil{2}\CorrespondingAuthor{chenlq@zju.edu.cn},
        Lingyun Sun\affil{1}\affil{2},
        Tingting Zhou\affil{3}
	}

\SetAffiliation{1}{College of Computer Science and Technology, Zhejiang University, Hangzhou, China}
\SetAffiliation{2}{International Design Institute of Zhejiang University, Hangzhou, China }
\SetAffiliation{3}{Alibaba Group, Hangzhou, China }
%	Note: Luis and Maria are not real people.  Henry and Catherine have been dead for >450 years.

%	To switch from inline author names to gridded names, use the [grid] option.

\maketitle

%%% Use this footnote for tracking various versions of your draft. Change text to suit your own needs. 
%%% \date{..} calls the same command. 
% \versionfootnote{Documentation for \texttt{asmeconf.cls}: Version~\versionno, \today.}% <=== Delete before final submission.

%%% Change these to your keywords.  Keywords are automatically printed at the end of the abstract.
%%% This command MUST COME BEFORE the end of the abstract.
%%% If you don't want keywords, leave the argument of \keywords{} empty (or use the abstract* environment)


%%%%%  End of fields to be completed. Now write your paper. %%%%%%%%%%%%%%%%%%%%%%%%%%%%%%%%%%%%%%%%%%%

%%%%%  ABSTRACT  %%%%%%%%%%%%%%%%%%%%%%%%%%%%%%%%%%%%%%%%%%%%%%%%%%%
%%
%% Abstract should be 200 words or less
\begin{abstract}
 UI-to-code technology has streamlined the front-end development process, reducing repetitive tasks for engineers. prior research mainly use design prototypes as inputs, with the effectiveness of the generated code heavily dependent on these prototypes' quality, leading to compromised robustness. Moreover, these approaches also exhibit shortcomings in code quality, including issues such as disorganized UI structures and the inability to support responsive layouts. To address these challenges, we introduce Prototype2Code, which achieves end-to-end front-end code generation with business demands. For Prototype2Code, we incorporate design linting into the workflow, addressing the detection of fragmented elements and perceptual groups, enhancing the robustness of the generated outcomes. By optimizing the hierarchical structure and intelligently recognizing UI element types, Prototype2Code generates code that is more readable and structurally clearer. To meet responsive design requirements, Prototype2Code primarily supports flexbox layout model, ensuring code compatibility across various device sizes. To validate the efficacy, we compare Prototype2Code with the commercial code generation platform CodeFun and Screenshot-to-code based on GPT-4 with vision. Employing structural similarity index measure (SSIM), peak signal-to-noise ratio (PSNR), and mean squared error (MSE) for visual similarity assessment, Prototype2Code's rendered UI effects align most closely with the design prototypes, exhibiting the minimal errors. We also conduct a user study with five experienced front-end engineers, inviting them to review and revise code generated by the three methods. As a result, Prototype2Code surpasses other methods in readability, usability, and maintainability, better meeting the business needs of industrial development.
\end{abstract}

\section{Introduction}

GUI-to-code generation has impacted the front-end industry, making development faster and cutting down on the repetitive work engineers encounter \cite{chen2018ui}. Leveraging deep learning advancements and extensive open-source datasets \cite{deka2017rico,li2022learning}, significant efforts are focused on developing techniques from visual representations, encompassing a fidelity from sketches \cite{mohian2020doodle2app}, wireframes \cite{feng2023designing} to actual UI screenshots \cite{chen2019automated}. However, the stringent requirements of large-scale industrial development, including adherence to specific frameworks and the need for maintainable, accessible code \cite{figma2code24}, typically lead to a process that starts with the creation of design documents. To this end, platforms like Imgcook \cite{imagecook} and CodeFun \cite{codefun} are utilized, serving as enterprise-level code generation platforms that cater to these stringent requirements. 

The input design documents, also referred to as design prototypes, are crafted by UI designers utilizing professional software such as Sketch \cite{sketch} and Figma \cite{figma}. Beyond visual representation, design prototypes incorporate a multimodal view hierarchy characterized by a tree-shaped layering structure, where each layer is endowed with attributes such as size, position, and type, thereby meeting the precise demands of development. However, the robustness of existing methods is often lacking, especially when designers do not consistently adhere to specifications, leading to lower quality designs. Such discrepancies in design quality, such as issues with fragmented elements and perceptual grouping, directly contribute to problems in the generated code. Existing solutions \cite{UILM,ULDGNN2022,chen2024egfe,xiao2022ui,xiao2024ui}, while primarily focusing on design linting to influence the generation process at the input stage, exhibit notable limitations. On one hand, their effectiveness is constrained by the dependency on specific generation algorithms, limiting their compatibility and adaptability across different platforms. On the other hand, without addressing the complexities inherent in the generation process, they often result in visual discrepancies and structural errors in generated UI effects, requiring further manual adjustments for accuracy. 

\begin{figure*}[tbh]
    \centering
    \includegraphics[width=0.9\textwidth]{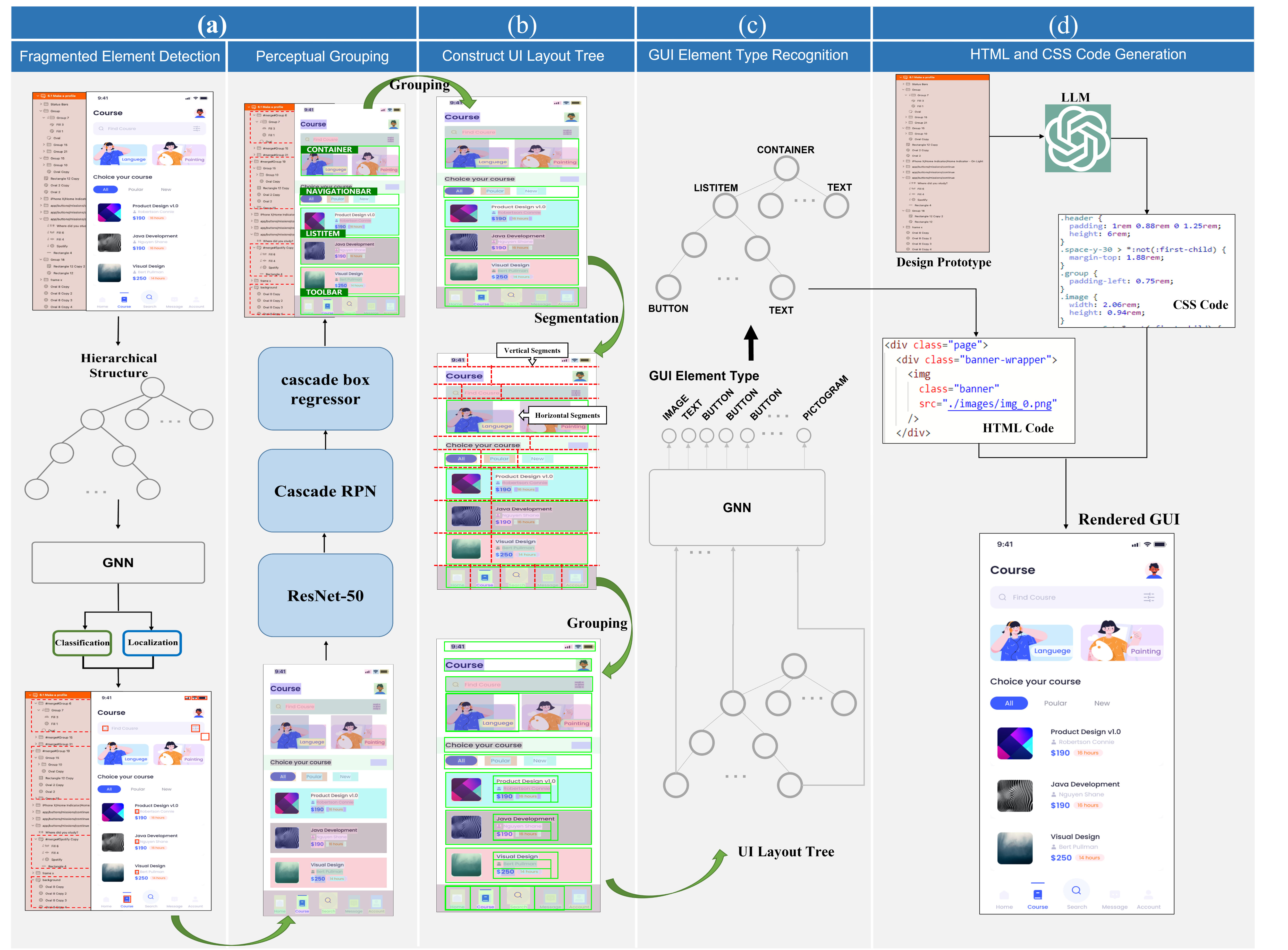}
    \caption{Overview of Our Proposed Prototype to Code Approach}
    \vspace{-0.11in}
    \label{fig:pipeline}
\end{figure*}

To address these limitations, we introduce Prototype2Code, an end-to-end solution that can automatically generate code from UI prototypes. Notably, we integrate the prototype linting (Fig. \ref{fig:pipeline} a) as a crucial step in the generation process, with the goal of elevating design quality and, in turn, bolstering the robustness of the code produced. Following this, the insights gained during the linting process are leveraged to guide the construction of the UI layout tree (Fig. \ref{fig:pipeline} b), ensuring that modifications made to the prototypes are accurately reflected in the final code structure. Furthermore, in constructing the UI layout tree, we take into account the industrial demand for responsive design, ensuring that the generated code is adaptable across devices of various sizes. By employing a graph neural network, we then determine the type of each UI element, such as text, image or list item (Fig. \ref{fig:pipeline} c). Finally, in the generation stage, we utilize the layout tree and recognized element types to construct the HTML skeleton code and leverage large language model for generating CSS style code, thereby rendering the final UI effect (Fig. \ref{fig:pipeline} d). To evaluate the effectiveness of our method, we conducted comparisons with results from generation platforms and other advanced approaches. Through quantitative visual similarity metrics, the UI effects rendered by our approach achieved the highest fidelity in reproducing the design intent. Our user study indicated that UIs generated by our method are more favored, and surpass other methods in terms of code readability, maintainability, and availability.

\section{Related Work}

\subsection{Code Generation Based on UI Prototypes}
In large-scale frontend industrial development, collaboration between designers and engineers is essential. Designers produce UI prototypes, which frontend engineers then utilize as a basis for coding. Due to the prevalence of numerous similar elements and functionalities in frontend applications, engineers often spend significant time on writing repetitive code \cite{chen2018ui, chen2024egfe}. To enhance efficiency, automatic code generation technologies for frontend have emerged. Compared to code generation based solely on visual representations, the path from design prototypes to code, enriched with multimodal information, better aligns with the requirements of industrial development \cite{figma2code24,chen2024egfe,ULDGNN2022}. 

It should be noted that, due to the need for design prototypes to be created by professional designers, obtaining them is much more challenging than acquiring visual representations. Most prototype to code methods originate from professional design platforms or large internet companies. For example, Figma to Code \cite{figmatocode} and Builder.io \cite{Builder} utilize engineering heuristic methods to parse prototype metadata for code generation. However, due to the lack of support for intelligent recognition of UI component types, their code fails to meet development specifications \cite{figma2code24}. Yotako \cite{yotako}, powered by WordPress, primarily relies on templates to create personal blogs and sharing websites. Other platforms, such as Imgcook \cite{imagecook} and CodeFun \cite{codefun}, employ deep visual techniques to recognize component and layout structures, converting the outcomes into a generic domain-specific language (DSL). Subsequently, they apply distinct rules to generate frontend code with various frameworks. Despite the efficiency in meeting large-scale demands, the process is hindered by ongoing challenges such as UI element fragmentation and perceptual grouping, necessitating further manual adjustments to the generated code. With large language models (LLMs) demonstrating exceptional performance in understanding requirements and generating code \cite{ikumapayi2023automated,nam2023ide}, platforms like Anima \cite{anima} have integrated generative AI technology, enabling the creation of code through requirements expressed in natural language. However, the performance of these LLMs in layout inference still falls short of expectations, with the UI effects they generate not matching the design drafts accurately.

\subsection{UI Fragmented Elements and Layers Grouping Problem}

In design prototypes, designers construct a hierarchical view structure by stacking layers and form the overall UI effects. Take Sketch for instance, it supports layers of diverse attributes, including vector, text, bitmap, and shape. Also, in practice, not all layers convey tangible elements; some serve as organizational containers, integrating several layers with distinct properties to form cohesive groups. The issue of fragmented elements was first identified by Chen et al. \cite{UILM}. A fragmented element group is typically composed of several basic vector or shape layers that together form a complete UI element, such as an icon. In the generated frontend code, these elements are not consolidated into a single icon image but exist as fragmented part instead, leading to redundant code structure and poor availability. They proposed a fragmented elements group detection method, using object detection techniques to locate potential groups and applying labels to the corresponding layers in the design prototypes to enforce consolidation. Li et al. \cite{ULDGNN2022} further clarified the concept of fragmented elements, defining them as layers that can not convey visual semantics independently. They primarily addressed the detection failures in previous visual methods caused by visual overlap among elements, such as the foreground and background occlusions in UI designs. Subsequent work, EGFE \cite{chen2024egfe}, extended their ideas by flattening UI elements in design prototype as a sequence and utilizing a Transformer to predict whether specific elements belong to fragmented elements. 

The issue of layers grouping, as another focal point, revolves around discussing how to achieve UI grouping in a nested way, thereby enhancing the modularity of the generated frontend code and facilitating partial reuse. UIGDL \cite{xiao2022ui}, Screen Recognition \cite{Recognition} and UISCGD \cite{xiao2024ui} addressed the identification of component-level groups, employing visual methods to locate groups containing the simplest UI functional units, such as a button comprising an icon and text. Inspired by the Gestalt principles, Xie et al. \cite{Xie2022PsychologicallyInspiredUI} further discussed the inference of section-level perceptual grouping, such as cards, lists, multi-tabs, and menus.

\section{Methodology}

\subsection{Prototype Linting}

Given a UI design prototype, its hierarchical structure unfolds in a tree-like manner, where each leaf layer node represents a specific UI element, such as text and image. Any non-leaf node encompasses multiple UI elements, forming specific perceptual areas and functional zones on the UI, such as lists and navigation bars. UI designs that do not strictly adhere to design standards may lead to deviations in this structure. In the issue of fragmented elements, a leaf node representing a fragmented element often fails to represent a UI element with complete semantics. For instance, a WiFi icon's three arcs-shape vectors might be represented by three separate leaf nodes. Regarding the perceptual grouping issue, some non-leaf nodes that indicate structure may be missing, causing the overall UI's nested structure to become less apparent. In the prototype linting section, we aim to achieve two objectives. First, we identify and mark fragmented elements that need to be merged; second, we perform perceptual grouping on the UI page to delineate regions. These information obtained is utilized in the second step to help construct the UI layout tree. 

In this paper, we adopted the approach developed by Li et al. \cite{ULDGNN2022} to identify fragmented layers. Compared to other methods, Li et al.'s approach has stood out for its superior efficacy, particularly in surmounting the hurdles of inference in cases of UI element occlusion \cite{UILM}, while also remaining resilient against the pitfalls of erroneous data patterns caused by uneven distribution of layer types \cite{chen2024egfe}. As illustrated in Fig. \ref{fig:pipeline} a, the reconstruction of the hierarchical structure was initially undertaken: the original layering structure present in the design prototypes was abandoned in favor of reevaluating containment relationships, which were determined solely based on the position and size (\(x,y,w,h\)) of each layer. Consequently, any parent node in the new layout tree was designed to physically enclose all elements of its children in the UI. 

Following this, we utilized a Graph Neural Network (GNN) for layer classification and bounding box regression. The classification branch was responsible for predicting whether a node belonged to a fragmented element, while the regression branch aimed at estimating the bounding box for the complete UI object associated with the current fragmented element. In the final step, groups of fragmented layers constituting a semantic UI element were marked. A \texttt{merge} attribute was added to their smallest parent node to indicate that these layers should be recognized as a single entity in subsequent processes. It is noted that all code and model weights from Li et al. were utilized, with all settings kept unchanged.

Following the identification of fragmented elements, we tackled the challenge of perceptual grouping of UI elements. This approach facilitated the creation of more structured and comprehensible code, thereby enhancing the performance and scalability of the generated front-end code in a more nuanced manner. Based on the CLAY dataset \cite{li2022learning}, we identified five types indicative of perceptual partitions: toolbars, navigation bars, cards, list items, and containers. The CLAY dataset, derived from the authoritative and widely used RICO dataset \cite{deka2017rico} in the UI field, has been optimized by removing noisy samples, thereby improving the performance of our tasks. By filtering the CLAY dataset, we obtained a total of 50,524 pages and 318,631 objects. The dataset was divided into training, test, and validation sets in a 7:2:1 ratio.

\begin{figure}[tbh]
    \centering
    \includegraphics[width=0.46\textwidth]{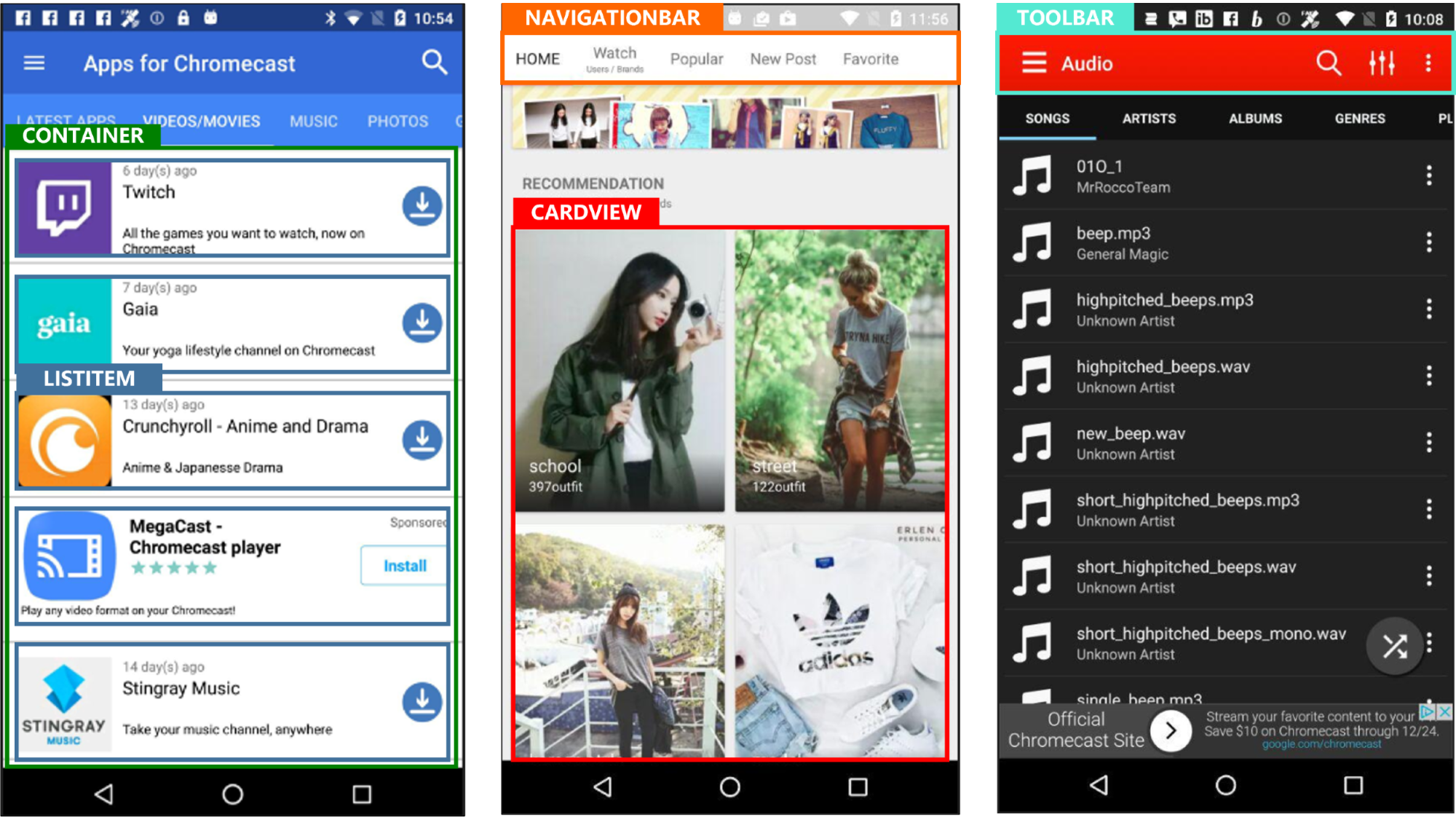}
    \caption{The Example of Perceptual Grouping Results}
    \vspace{-0.11in}
    \label{fig:group_detection}
\end{figure}

Given that visual detection methods have been extensively validated for their effectiveness in UI scenarios, we adopted the SOTA detection method proposed by Chen et al. \cite{UILM} to facilitate the inference of perceptual groups. They employed ResNet-50 \cite{he2016deep} as the backbone to process UI visual representation, integrating Cascade RPN and a cascade box regressor \cite{vu2019cascade} to incrementally refine the bounding boxes. To enhance detection performance, they introduced the boundary information of UI elements as additional knowledge to guide the inference process. We constructed the model based on the MMDetection \cite{chen2019mmdetection} framework and trained it using the filtered CLAY dataset, with all parameter settings being identical to those reported by Chen et al. in UILM. As shown in Fig. \ref{fig:group_detection}, following this step, we have segmented the UI interface according to perceptual zones, providing guidance for the structure of frontend code, which is conducive to modular expression. 

\subsection{Constructing the UI Layout Tree}
\label{UI_layout_tree}
In this section, we explore the construction of the UI layout tree from design prototypes. Our approach is guided by two primary objectives: firstly, to address issues related to fragmented layers and the loss of perceptual groups; secondly, to ensure that the revised structure accommodates the demands of responsive layouts.

For all layers in the design prototypes, we start by reconstructing their containment relationships in the prototype based on the detection results of perceptual groups. We arrange all perceptual groups in ascending order of size and iterate them through all layers. This strategy is adopted to manage potential nesting issues within the UI, where larger perceptual groups may include multiple smaller ones. If the Intersection over Union (IoU) ratio between the current layer and a given perceptual group exceeds a predefined threshold, we consider the UI element represented by that layer to belong to the current group. Fragmented layers tagged with the \texttt{merge} attribute are treated as a single element, with their position and area calculated from the smallest top-left point to the largest bottom-right point coordinates. For UI elements that belong to the same perceptual group, we add a \texttt{group} label to their parent layer. 

\begin{figure}[tbh]
    \centering
    \includegraphics[width=0.48\textwidth]{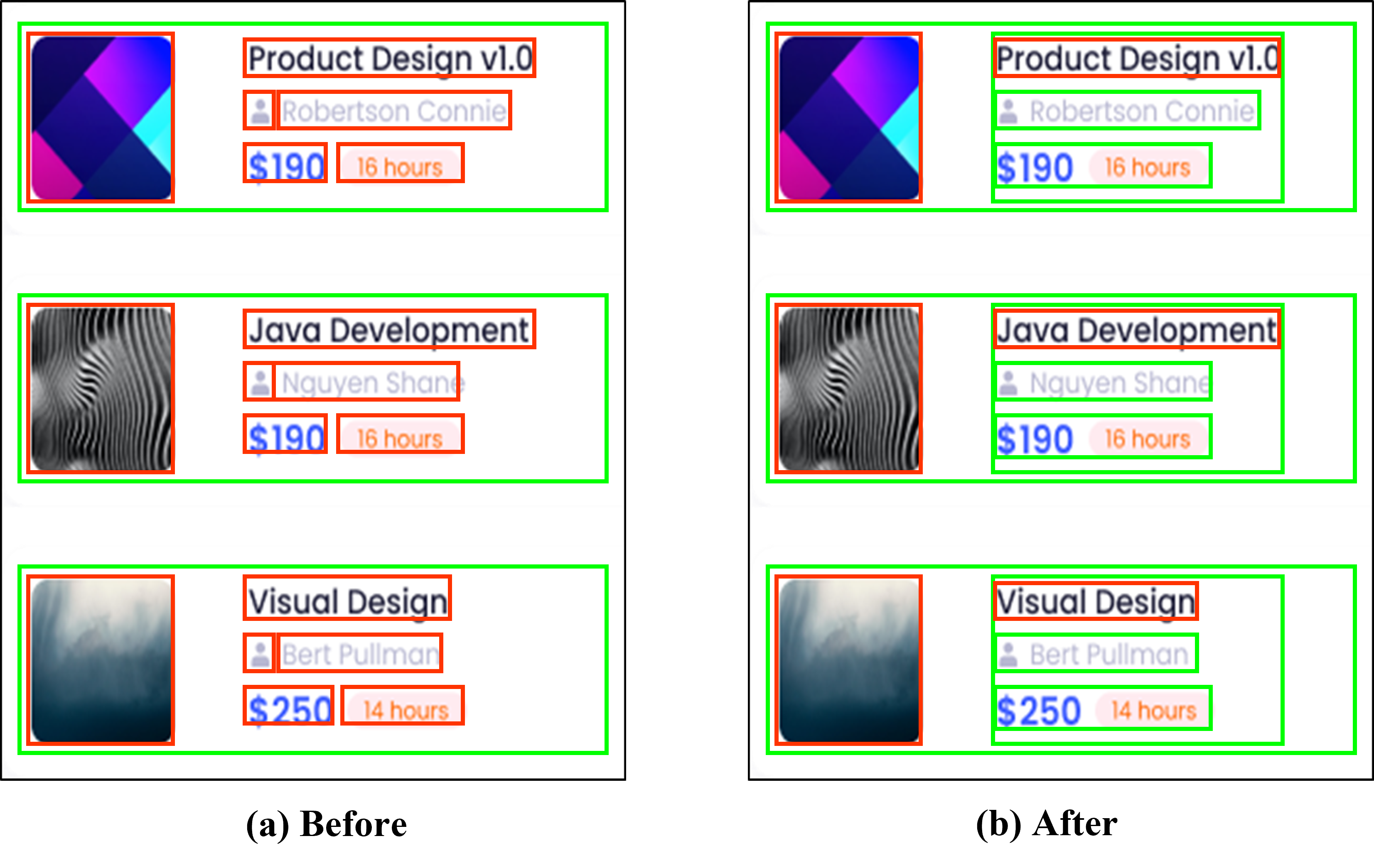}
    \caption{The Layout Structure Before and After Row/Column Re-segmentation}
    % \vspace{-0.11in}
    \label{fig:layout}
\end{figure}

We then create the UI layout tree based on the optimized hierarchical structure. As illustrated in Algorithm~ \ref{layout_construct}, given a set of layers with hierarchical structure ${layer_i}_{i=1}^{N}$, we aim to output a tree structure $T$ that consists of a set of nodes $V$ and a node relation matrix $M\in \mathbb{R}^{N\times N}$, $M[i,j] = 1$ indicates that node $i$ is the parent node of node $j$. Our entire algorithm is divided into two phases: layout tree initialization and row/column re-segmentation for flexbox layouts. For the layout tree initialization, we extract all leaf layers representing UI elements as the set $E$. For those belonging to fragmented layers or part of perceptual groups, we take out the element represented by their parent node as a whole representation. For all the UI elements obtained in $E$, we sort them in descending order by their size and iterate through them, adding the current element to the tree under the node with which it has a containment relationship and the smallest size. Following this, we deal with the overlapping issue among UI elements, which affects the generation of layout-related CSS code. For each non-leaf node in $T$, we check its subordinate nodes for overlaps and generate new parent nodes for them with overlapping, subsequently attaching it to their original parent node. We also add an attribute \texttt{need\_absolute=True} to the newly generated parent nodes, indicating that the elements under these nodes require absolute positioning in their layout. 
\begin{algorithm}[tbh]
    \caption{UI Layout Tree Construction Algorithm}
    \label{layout_construct}
    \begin{algorithmic}[1] % 有序号显示
    \Require the prototype structure $\{layer_i\}_{i=1}^{N}$
    \Ensure the UI layout tree $T=\{V,M\}$
    \Statex Phase 1: Layout tree initialization
    \State $E = T.LeafLayers().map(L \rightarrow$
    \Statex \quad $L.isFrag$ || $L.isInGroup$ ? $L.parent() : L)$
    \For{$e$ in $E.descending()$}
        \State $V = V \cup \{e\}$
        \State $M[e.findparent(T), e.index()] = 1$
    \EndFor
    \For{$e$ in $T.NonLeafLayers()$}
        \For{$O = e.Overlap().Empty() ? e.Overlap() : \{\}$}
            \For{each element set $S$ in $O$}
                \State $V = V \cup \{NewNode(need\_absolute=True)\}$
                \State $M[V.len(), i] = 1$ for $i$ in $S$
            \EndFor
        \EndFor
    \EndFor
    \Statex Phase 2: Row/column re-segmentation
    \Repeat
        \State Create vertical segments based on $Y$ coordinates
        \State Create horizontal segments based on $X$ coordinates
    \Until {No more segments can be added}
    \For{each segment $S$ in vertical segments}
        \State $V = V \cup \{NewNode(flex-direction=row)\}$
        \State $M[V.len(), i] = 1$ for $i$ in $S$
    \EndFor
    \For{each segment $S$ in horizontal segments}
        \State $V = V \cup \{NewNode(flex-direction=column)\}$
        \State $M[V.len(), i] = 1$ for $i$ in $S$
    \EndFor
    \State \Return the UI layout tree $T=\{V,M\}$
    \end{algorithmic}
\end{algorithm}

% Fig. \ref{fig:layout} a demonstrates a structural effect up to this point. The red boxes represent UI elements as leaf nodes, while the green boxes are their parent nodes as groups.

In order to achieve a responsive design, we utilize flexbox layouts for UI elements. This requires that UI elements under a parent node have a uniquely determined arrangement and their order can be defined. Fig. \ref{fig:layout} a demonstrates a structural effect up to this point. Within the green-boxed parent node, elements in red boxes are inconsistently arranged, some vertically and others horizontally. Therefore, in the second phase of Algorithm~\ref{layout_construct}, we conduct row/column re-segmentation to further refine the groups, ensuring that elements within a group are explicitly ordered either by rows (left to right) or by columns (top to bottom). Generally, this involves understanding the relative positions of elements on a UI page, which has been widely discussed in the field of UI accessibility. In this case, we enhanced the segmentation algorithm from Screen Recognition \cite{Recognition}, which is based on the reading order from top to bottom and left to right, to determine the order of elements on a page as we required. In summary, the process is divided into three steps: (i) We create vertical segments by identifying $Y$ coordinates at which a horizontal line can be drawn across the page without intersecting any elements. (ii) Similarly, within these vertical segments, we form horizontal segments using $X$ coordinates, following the same non-intersecting principle. (iii) This first two steps are recursively applied to all horizontal segments until no more segments can be added. Then, based on the segmentation results, we determine the horizontal layout among UI elements. If multiple UI elements are within the same horizontal segment, we group them together under a new parent node, setting its flex layout property to \texttt{flex-direction=row}, as illustrated by the elements ``\$190'' and ``16 hours''. Subsequently, we examine the vertical layout. In cases where multiple UI elements are in the same vertical segment, we group them and set the flex layout property to \texttt{flex-direction=column}. The final effect (Fig. \ref{fig:layout} b) allows us to ascertain the arrangement and order of any element relative to its parent node.

\subsection{UI Element Recognition}

\begin{figure}[tbh]
    \centering
    \includegraphics[width=0.47\textwidth]{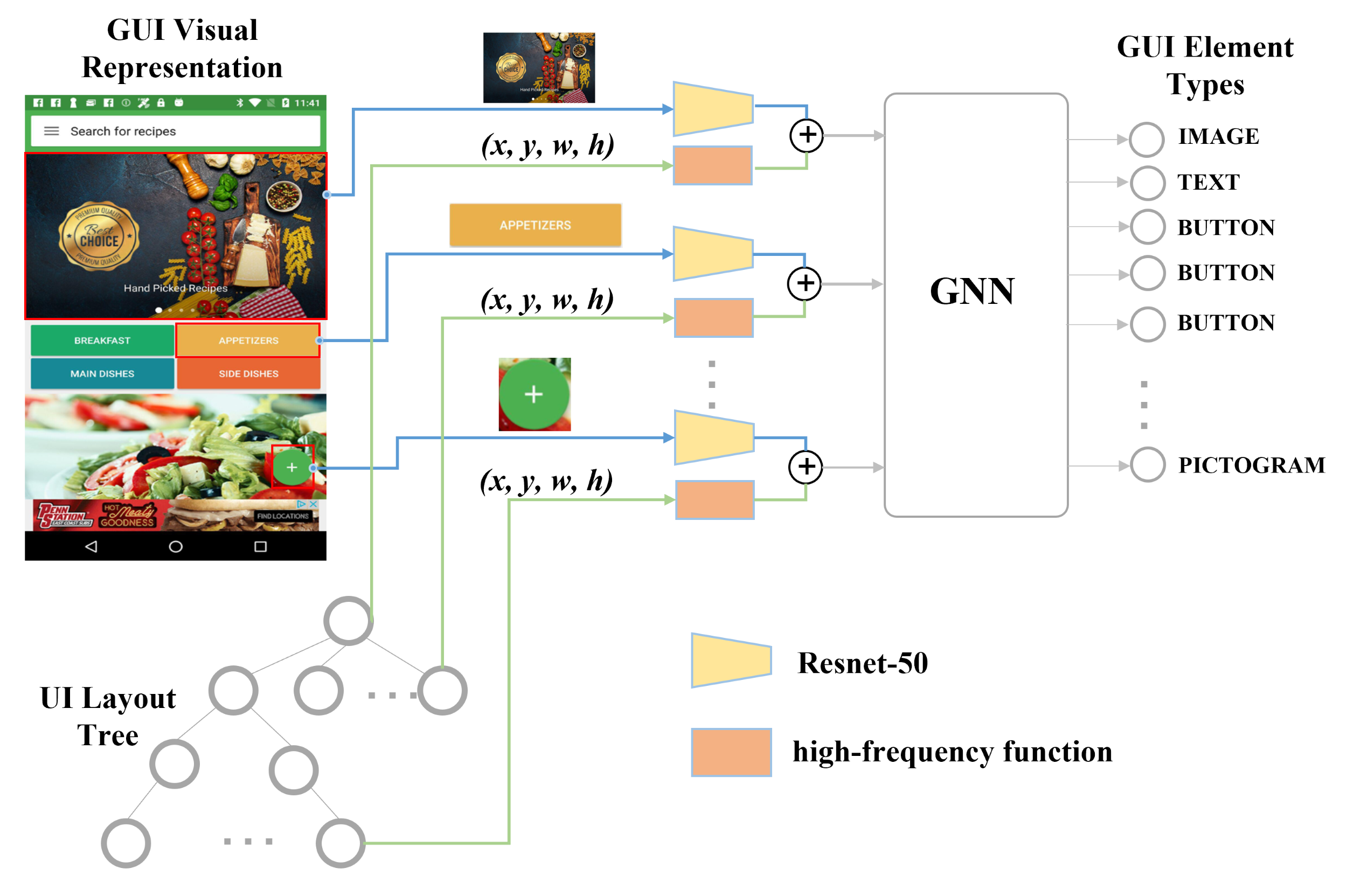}
    \caption{Our Approach to Predict UI Element Types}
    % \vspace{-0.11in}
    \label{fig:type_pred}
\end{figure}

So far, we have laid the groundwork for generating the HTML code. In this section, we explore the process of assigning specific element types to each UI element within the UI layout tree. These types, such as ``icon'', ``status-bar'' and ``button'', are utilized as class names for each \texttt{<div>} element. By implementing this approach, we enrich the semantics of the HTML code, thereby making it more readable and promoting the reuse of specific components \cite{figma2code24}. 

Figure \ref{fig:type_pred} presents our approach for the recognition of UI element. Reflecting on the insights from prior research \cite{chen2024egfe,chen2020object,zang2021multimodal}, which highlighted the value of multimodal information in enhancing UI element identification, our method integrates both the visual representation and spatial dimensions (\(x,y,w,h\)) of each element. For each node in our UI layout tree, we extract its visual representation, scale it to a size of 64x64 pixels, and employ ResNet-50 to obtain image feature $I$. Regarding the spatial dimensions information, we encode it into a high-dimensional vector $S\in \mathbb{R}^{L}$ using the following high-frequency function:
\begin{equation}
\scriptstyle
\gamma(z) = \left( \sin(2^0 \pi z), \cos(2^0 \pi z), \ldots, \sin(2^{L-1} \pi z), \cos(2^{L-1} \pi z) \right).
\end{equation}

Here, $\gamma(z)$ maps the position and size information of each UI element one-to-one into a higher-dimensional feature space, facilitating better integration with other features during subsequent forward propagation. We embed both the image feature $I$ and the spatial feature $S$ into 128-dimensional vectors, and apply element-wise addition to combine them. This sum serves as the initial state information for each node within the graph neural network(GNN). 

To address the limitations of GNN such as oversmoothing and oversquashing, we draw inspiration from prior work \cite{ying2021transformers, rampavsek2022recipe} and incorporate a multi-head self-attention mechanism into our model as shown in Equation~\ref{selfatt}. Specifically, $X^{L+1}$, representing the state of the nodes at the $(L+1)^{th}$ layer of the network, is composed of two parts: $X^{L+1}_M$ originated from the graph network itself and $X^{L+1}_T$ derived from the multi-head attention mechanism. 
\begin{equation}
\label{selfatt}
\scriptstyle
\begin{aligned}
X^{L+1} &= \text{MLP}^L(X^{L+1}_T + X^{L+1}_M), \\
X^{L+1}_T &= \text{Self-attn}^L(X^L), \\
X^{L+1}_M &= \text{MPNN}^L(X^L), \\
X^{L+1}_M &= \text{LayerNorm}(\text{Dropout}(X^{L+1}_M) + X^L), \\
X^{L+1}_T &= \text{LayerNorm}(\text{Dropout}(X^{L+1}_T) + X^L). \\
\end{aligned}
\end{equation}
In this context, MLP refers to a multi-layer perceptron, a neural network module where each layer is fully connected to the subsequent one. The self-attention module, crucial in Transformer architectures, assigns weights to different parts of the input, enhancing focus on relevant features. The MPNN (Message Passing Neural Network), facilitates effective graph learning by enabling the exchange of messages among nodes via their connecting edges \cite{gilmer2017neural}. LayerNorm stands for layer normalization, and dropout is employed to prevent overfitting in the learning process. 

After $K$ rounds of GNN state updates, we employ a two-layer MLP to accomplish the recognition of element types. We trained the model on the CLAY dataset \cite{li2022learning}, comprising 59,555 UI pages with 1,368,383 objects across 25 types. The dataset was split into a 7:2:1 ratio for training, testing and validation. The initial learning rate was set at 
1e-4, with a decay rate of 0.99. Due to the imbalance in UI element types in dataset, we adjusted the weights of the cross-entropy loss according to the frequency of types in the training set, setting the weight to be the inverse of the frequency.

\subsection{HTML and CSS Code Generation}
Finally, in this section, we introduce the method for generating UI front-end code, which is primarily divided into two parts: the HTML skeleton and the CSS style code. With the development of multimodal large language models(LLMs) in the field of code generation, recent studies have explored their performance in front-end code generation. Among these, the open-source tool Screenshot-to-Code \cite{s2c} has emerged as one of the most popular and representative one. Despite its capabilities, Screenshot-to-code still lacks the ability to parse input GUIs accurately, limiting it to generating rendered code that only stylistically matches the design input. Learning from these insights, we decided to separate the tasks of layout generation (HTML) and style generation (CSS) (Fig. \ref{fig:pipeline} d). This approach leverages the strengths of LLMs in text generation for CSS code production, while also maintaining the UI layout tree's strong guidance for the HTML code structure.

We first demonstrate how to derive HTML code from the UI layout tree. For any non-leaf node, we encapsulate it with a \texttt{<div>} tag and traverse the subtree rooted at it, using the nesting of \texttt{<div>} tags to depict the entire skeleton layout. For leaf node elements representing images and text, we apply \texttt{<img>} and \texttt{<span>} tags, respectively. Utilizing the results of UI element category recognition, we assign a classname to each element to serve as a CSS selector.

CSS styles dictate the rendering effects of elements on a page. There are two types of CSS styles: layout-related properties determine the rendering position of elements on the page; visual effect-related properties, on the other hand, ensure the UI's aesthetic appeal. Previous research \cite{codefun,s2c} have predominantly supported absolute size layouts, leading to pages that cannot respond to device sizes, which is not conducive to further development and maintenance. With this consideration, we adopt a flex layout model that can respond to device sizes. For each element wrapped in a \texttt{<div>}, we determine the layout method based on the attributes set in Section~\ref{UI_layout_tree}. For nodes with the \texttt{need\_absolute} attribute, we set their CSS \texttt{position} property to \texttt{relative}, and their children's \texttt{position} property to \texttt{absolute}. Here, child nodes are absolutely positioned based on the top-left corner coordinates of the parent node, setting their \texttt{top} and \texttt{left} properties. If a child node fits the characteristics of row or column arrangement, we set the parent node's \texttt{position} property to \texttt{flex}, and arrange it in rows or columns according to the \texttt{flex-direction} property determined in Section~\ref{UI_layout_tree}. The  \texttt{margin-top} and \texttt{margin-left} properties of each child node are determined by their distance from the parent node or adjacent nodes. The \texttt{width} and \texttt{height} properties can be directly obtained based on parameters in design prototypes.

Finally, we describe how to utilize LLMs such as GPT-4 \cite{OpenAIChat2023} for generating the remaining visual effect-related CSS properties. Our prompt comprises four parts: <Role-playing, User Input, Field Explanation, Output Requirement>.  In the role-playing part, we prompt the LLM as a front-end developer with years of experience and advanced technical skills. For user input, based on the Figma API user documentation \cite{figmaapi}, we filter out style information unrelated to visual effects, organizing the remaining information in a dictionary key-value pair format. In the field explanation part, we review the Figma API documentation again, storing interpretative texts for all relevant style properties as a JSON file, enabling the LLM to understand the style in the design prototypes. In the output requirement, we instruct the LLM to return a parsable JSON, where each field corresponds to a related CSS property and its value.

\section{Experiments And Results}

\subsection{Perceptual Grouping Performance}

First, we assess the performance of our perceptual group detector. The advantage of the cascading network lies in its ability to refine bounding boxes through multi-stage regression, reducing the noise from low-quality candidate boxes while also preventing overfitting. As shown in Table~\ref{tab:Detection}, we report the object detection performance using the standard COCO metrics \cite{lin2014microsoft}. Considering the significant size variations among UI elements, we also provide separate Average Precision (AP) scores for large ($>96^2$ pixels), medium ($32^2-96^2$ pixels), and small ($<32^2$ pixels) targets. 

For comparison, we include three major types of object detection models as baselines. The Faster-RCNN \cite{ren2015faster} and its optimized variant, GA-Faster-RCNN \cite{wang2019region}, represent the two-stage anchor-based bounding box refinement approach (Type 1). CenterNet \cite{duan2019centernet} and YOLOF \cite{chen2021you}, which are anchor-free and operate through end-to-end bounding box regression (Type 2), are similar in model size to our detector and have demonstrated commendable performance on the COCO dataset. Lastly, Deformable-DETR \cite{zhu2020deformable}, which is built upon the Transformer architecture (Type 3), serves as a quintessential example of this type in the object detection framework.

\begin{table}[htp]
\centering
\caption{Results of Perceptual Group Detection (IoU 0.50:0.95)}
\vspace{-0.04in}
\label{tab:Detection}
  \begin{tabular}{lcccc}
    \toprule
    \textbf{Method} &\textbf{$mAP$} &\textbf{$AP_s$} &\textbf{$AP_m$} & \textbf{$AP_l$}\\
    \midrule
    Faster-RCNN & 0.583 & 0.197 & 0.287 & 0.597  \\
    GA-Faster-RCNN & 0.617 & 0.200 & 0.306 & 0.629   \\
    CenterNet & 0.567 & 0.179 & 0.297 & 0.576  \\
    YOLOF & 0.617 & 0.107 & 0.266 & 0.633   \\
    Deformable-DETR &0.640 & 0.126 & 0.279 & \textbf{0.655}  \\
    Ours &\textbf{0.640} & \textbf{0.219} & \textbf{0.369} & 0.651   \\
    \bottomrule
  \end{tabular}
  % \vspace{-0.11in}
\end{table}

As a result, our detector, categorized within Type 1, surpassing both Faster-RCNN and GA-Faster-RCNN with an increase of 5.7\% and 2.3\% in mean average precision (mAP), and approximately a 2\% improvement in precision for detecting small targets. Among the second type of models, CenterNet performs less effectively, trailing our detector by 7.3\% in mAP. The YOLOF model shows better results but falls short in precision for small and medium elements. Similarly, Deformable-DETR, as the third type of model, show less precision in grouping small elements, with an 11.2\% lower performance compared to our detector. Overall, its mAP is roughly on par with our model, and it slightly surpasses our model by 0.4\% in recognizing large targets. However, Deformable-DETR requires more VRAM and longer training times in terms of training resources.

\subsection{Element Type Recognition Performance}

In this section, we report on the performance of UI element type recognition. Identifying the types of UI elements is a crucial step in the front-end code generation process, as it enhances the semantic richness of the code and improves its readability. As illustrated in Table~\ref{tab:recognition}, we evaluate the model using three metrics: 
\begin{equation}
\label{recog_metrics}
\scriptstyle
\begin{aligned}
precision &= \frac{TP}{TP+FP}, \\
recall &= \frac{TP}{TP+FN}, \\
F1~score &= 2 \times \frac{precision \times recall}{precision+recall}.\\
\end{aligned}
\end{equation}
Given the CLAY dataset's composition of 25 UI element types and its imbalanced sample distribution, we present not only the default macro average but also the results of the weighted average to provide a comprehensive performance overview.

In the task of recognizing UI element types, certain elements, like text elements and text buttons, exhibit high visual similarity, potentially causing neural networks to make incorrect type classifications. To address this issue, we employ a Graph Neural Network (GNN) to achieve a comprehensive feature representation for each element. Utilizing the message-passing capabilities of GNN allows us to capture the contextual and semantic relationships between elements within the GUI layout. This method significantly improves the uniqueness of the representation vectors, leading to enhanced accuracy in type classification. As a validation, we compare our method with a baseline model excluding GNN (line 1 in Table~\ref{tab:recognition}). The inclusion of GNN has led to noticeable improvements: a rise in the weighted average metrics—precision, recall, and F1 score by 4.3\%, 4.4\%, and 4.8\% respectively. While a slight decrement in macro average precision is observed, significant enhancements have been observed in recall and F1 score, each by around 5\%.

\begin{table}[htp]
\centering
\vspace{-0.11in}
\caption{Results of Element Type Recognition}
\label{tab:recognition}
\resizebox{0.48\textwidth}{!}{
 \begin{tabular}{lcccccc}
  \toprule
  & \multicolumn{3}{c}{\textbf{Weighted Average}} & \multicolumn{3}{c}{\textbf{Macro Average}} \\
  
  \midrule
  
  \textbf{Method} & \textbf{precision} & \textbf{recall} & \textbf{F1 score} & \textbf{precision} & \textbf{recall} & \textbf{F1 score}\\
  \midrule
 ResNet & 0.759 & 0.761 & 0.754 & \textbf{0.710} & 0.591 & 0.628 \\
 ResNet+GNN (ours) & \textbf{0.802} & \textbf{0.805} & \textbf{0.802} & 0.708 &\textbf{0.649}& \textbf{0.673} \\
 ResNet+RoI+GNN & 0.737 & 0.739 & 0.727 & 0.641 & 0.526 & 0.555 \\
 ResNet+FRNRoI+GNN & 0.733 & 0.737 & 0.723 & 0.669 & 0.525 & 0.568 \\
  \bottomrule
 \end{tabular}
 }
% \vspace{-0.11in}
\end{table}

Another issue that has garnered attention lies in the expression of elements' visual representation. In our method, we uniformly scale all UI elements to a size of 64x64 pixels which is a common practice applied. However, UI pages often contain elements that are either too wide or too narrow. For instance, a toolbar may have an aspect ratio exceeding 6. In such cases, uniform scaling can distort the elements, leading to a loss of semantic information and poor representation. To avoid information loss due to image scaling of components, previous methods have employed a strategy combining RoIPooling and RoIAlign \cite{ULDGNN2022,Magic_Layouts} to obtain feature maps that match the original dimensions of the elements. To validate the performance of this strategy, we introduce two additional baselines. The RoI method (line 3 in Table~\ref{tab:recognition}) extracts the feature map of the entire UI using ResNet-50, and then employs the RoIAlign algorithm to obtain representations for each single element region. For the FPNRoI method (line 4 in Table~\ref{tab:recognition}), it further enhances the FPNRoI method by utilizing a Feature Pyramid Network (FPN) to integrate high-dimensional and low-dimensional feature expressions of the UI page. From the results, the approach of uniformly scaling to extract elements features (our method) proves to be more suitable for UI element recognition. Compared to the suboptimal RoI method, it achieves improvements in the weighted average precision, recall, and F1 score by 6.5\%, 6.6\%, and 7.5\%. A possible reason may stem from the data-centric nature of the model: since ResNet-50 is pretrained on the ImageNet dataset, which does not contain UI images, the features it extracts may not effectively match those found in UI elements.

\begin{figure}[tbh]
    \centering
    \includegraphics[width=0.48\textwidth]{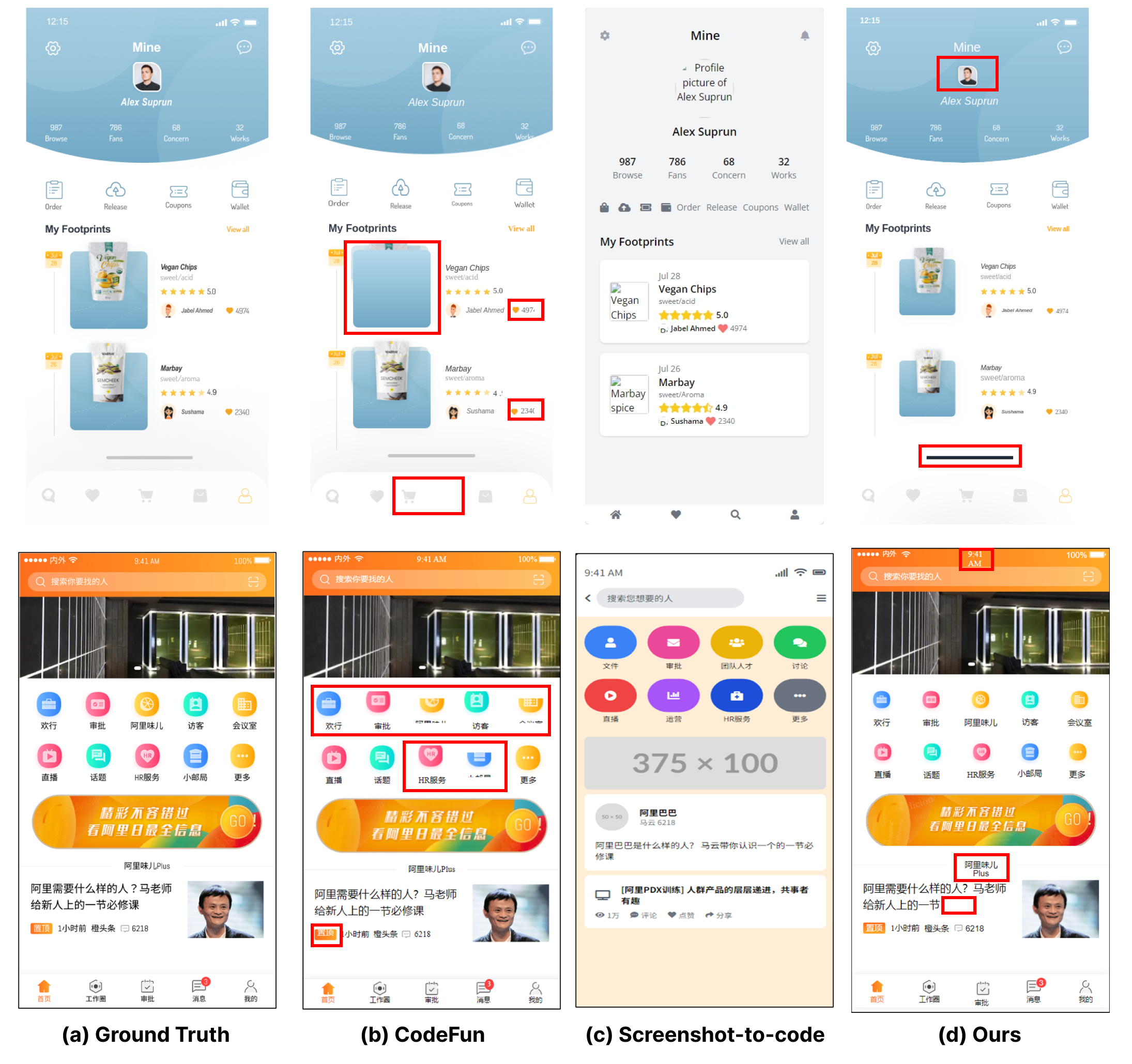}
    \caption{Comparison of Rendered Effects}
    \vspace{-0.11in}
    \label{fig:rendered_effect}
\end{figure}

\subsection{Visual Quality Assessment}

Assessing the alignment of rendered GUI visual effects with original design intentions is crucial for evaluating front-end code generation. In this section, as shown in Table~\ref{tab:visual_metrics}, we introduce three metrics based on visual similarity for the consistency evaluation. Among them, structural similarity index (SSIM) evaluates from the perspectives of luminance, contrast, and structure, yielding results that closely mirror the intuitive perceptions of human; peak signal-to-noise ratio (PSNR) and mean squared error (MSE) measure absolute errors, assessing image similarity based on the differences in corresponding pixel values. Higher SSIM and PSNR values indicate that the image quality is closer to the true value, while for MSE, a lower value signifies better consistency.

We select two baseline models for our comparison: CodeFun \cite{codefun} is a commercial code generation platform that supports large-scale industrial development while Screenshot-to-code \cite{s2c} is currently the most popular open-source generation tool, implemented based on GPT-4v. It is important to note that Screenshot-to-code only supports code generation from UI screenshots and, compared to our method and CodeFun, lacks the multimodal information contained in design prototypes, making the comparison somewhat unfair. However, considering that the use of multimodal LLMs to assist in front-end code generation may become a trend in the future, we include it for comparison and discuss the existing shortcomings. 

Regarding the dataset, we select 10 design prototypes from the fragmented elements dataset by Chen et al. \cite{UILM}, which contains a large number of design errors, representing outputs from relatively low-skilled designers. Additionally, we acquired another 20 free designs from the Figma community, each averaging over 200 likes, representing higher-quality designs. Statistically, we conducted validations on 30 design prototypes, encompassing a total of over 4000 layers.

\begin{table}[htp]
\centering
\caption{Results of Visual Quality Assessment}
\vspace{-0.04in}
\label{tab:visual_metrics}
\resizebox{0.48\textwidth}{!}{
  \begin{tabular}{lccc}
    \toprule
    & \textbf{SSIM $\uparrow$} & \textbf{PSNR $\uparrow$} & \textbf{MSE(1e-2)$\downarrow$} \\
    \midrule
    \textbf{Ours} &\textbf{0.91 ± 0.07}  &\textbf{21.15 ± 2.01}  &\textbf{0.97 ± 0.42}  \\
    \textbf{CodeFun} &0.64 ± 0.06  &17.42 ± 2.26  &  3.57 ± 1.45  \\
    \textbf{Screenshot-to-code} &0.58 ± 0.12   &0.58 ± 0.12   & 10.46 ± 2.35 \\
    \bottomrule
  \end{tabular}
}
% \vspace{-0.11in}
\end{table}

Quantitatively, the UI effects rendered by our method are significantly superior to those of CodeFun and Screenshot-to-code, with improvements of 0.27 and 0.33 in SSIM; 3.73 and 9.14 in PSNR; and a reduction of 2.6 and 9.49 in MSE, respectively. Qualitatively, we demonstrate the differences in two rendered effects as Fig. \ref{fig:rendered_effect}. We have marked the evident errors with red boxes. Comparing CodeFun with our method, the former exhibits more errors, with its image and text elements showing incomplete displays. In contrast, the discrepancies in our method primarily stem from differences in the applied CSS visual properties, which can be rectified by adjusting the target object. Unlike other methods, Screenshot-to-code does not achieve a faithful replication of designs; its code-rendered images bear only a stylistic resemblance to the original designs. Furthermore, due to the inherent randomness in its output, which results in variability with each generation, we believe it does not meet the requirements of industrial development so far.

\subsection{User Study on Code Quality}

\begin{figure}[tbh]
    \centering
    \includegraphics[width=0.46\textwidth]{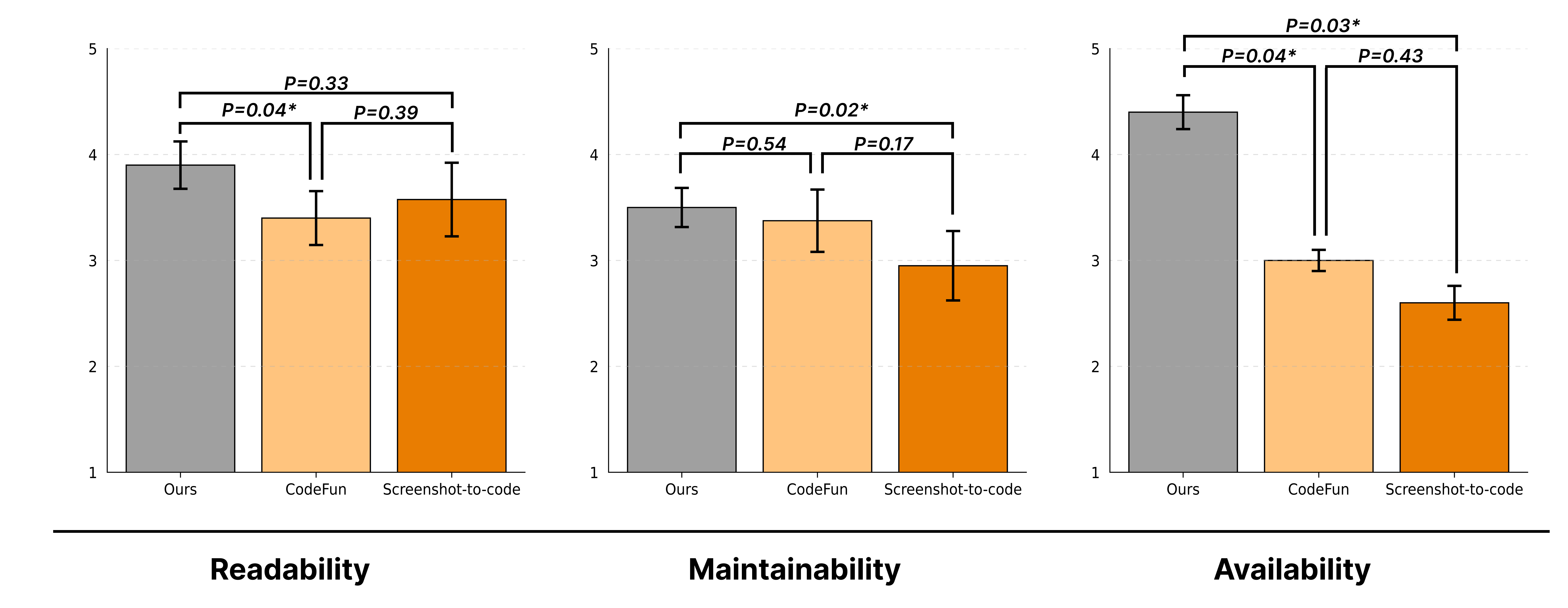}
    \caption{Results of User Study}
    % \vspace{-0.11in}
    \label{fig:user_study}
\end{figure}
Beyond evaluating the rendered visual effects, the overall quality of code is a crucial criterion for assessing the performance of automatic generation methods. In large-scale industrial development, code readability, availability, and maintainability are three key metrics for evaluation \cite{chen2024egfe,UILM}. Readability assesses whether the generated code is easy to understand, featuring a clear structure and using of meaningful names. Availability measures the extent to which the code requires manual modifications, calculated as: $1-\frac{lines\ of\ code\ changes}{total\ lines\ of\ code}$. Meanwhile, maintainability evaluates the difficulty of maintaining the code. 

For our user study, we invite five front-end engineers with an average work experience of over two years. Each participant is asked to review the code for 10 generated UI pages. For readability and maintainability, they report their assessments on a 5-point Likert scale. Regarding availablity, participants are required to modify the code for each page until it matches basic business requirements. We used git to track code modifications and, following the method of Chen et al. \cite{chen2024egfe}, mapped the original scores starting from 0.8, with each 5\% interval corresponding to a 1-5 scale. 

We present the average scores in Fig. \ref{fig:user_study}, with error bars indicating the standard error, and we report the p-values using the Mann-Whitney U test. For readability, our method achieves a score of 3.9, surpassing CodeFun's 3.4 and Screenshot-to-code's 3.58. Screenshot-to-code performs slightly better than CodeFun in terms of readability, primarily because it generates less code with a simpler structure. For the same reason, we did not find any significance in readability. In terms of maintainability, our method scores 3.5, also higher than CodeFun's 3.38 and Screenshot-to-code's 2.95. Significantly, a notable difference was observed between our method and Screenshot-to-code ($p=0.02^*$), with participants unanimously agreeing that our method performed better in responsive design compared to the others. No significant difference was observed between our method and CodeFun ($p=0.54$); a potential shortcoming of our method lies in its less effective reuse of certain container CSS properties compared to CodeFun. Finally, we observed significant differences in availability ($p=0.04^*$ with CodeFun and $p=0.03^*$ with Screenshot-to-code). Our method scores 4.4, significantly higher than CodeFun's 3.0 and Screenshot-to-code's 2.6. For our method, participants rarely needed to adjust the HTML structure and only had to make minor modifications to the CSS to meet business requirements.

\section{Discussion and Future Work}

In this paper, we aim to bridge the gap in generating front-end code from GUI design prototypes, implementing an end-to-end approach that encompasses five key steps: design prototypes linting, code layout tree construction, GUI element recognition, and code generation. Design prototypes linting is a critical step for automated code generation, especially for designs with quality deficiencies. Incorrect hierarchical structures \cite{UILM,chen2024egfe,ULDGNN2022} have been shown to adversely affect both the rendering effect and functionality of the generated code. In our approach, we primarily include detection of fragmented element layers and perceptual grouping processes. These two components together form an understanding of nested grouping within UI interfaces: starting from UI fragmented elements with incomplete semantics, moving to image-text groups with complementary semantics, and finally obtaining functional groups at a larger component scale. The UI structure generated on this basis is more organized and less prone to errors. The case2 in Fig. \ref{fig:rendered_effect} offers an example. In the effect rendered by CodeFun, we can see some image and text elements were incompletely displayed, resulting in positional offsets. This occurred because some graphic-text elements were paired and placed under a single ``div'', while others were wrapped in separate ``divs,'' leading to errors during rendering based on CSS properties. In our method, however, this part of the code contains three nested div layers: the innermost container wraps each image-text group, the second layer encloses a row of five icons, and the outermost layer contains two row elements. In such scenarios, CSS properties can be highly reusable and less prone to errors. For the fragmented elements detection module, as we directly used the same code and model weights from Li et al. \cite{ULDGNN2022}, we did not further validate its effectiveness in our experiments. As for perceptual grouping, the current overall recognition accuracy has only reached 0.64. Continuing to improve the detection accuracy will be a direction for future efforts. Some potential strategies include incorporating larger Transformer models trained with a significant number of parameters, applying more UI data for specialized training, or making fuller use of multimodal UI data.  

Through visual quality assessments, we compared the UI rendering effects produced by Prototype2Code with those of CodeFun and Screenshot-to-code. The results indicate that our method more faithfully reproduces design prototypes with fewer errors. From the perspective of large-scale development, both our method and CodeFun generally meet the needs for page effect realization. Screenshot-to-code, however, still has much room for improvement. Considering that the quality of the generated code cannot be fully assessed through visual effects alone, we recruited professional front-end engineers and sought their advice on the code's readability, usability, and maintainability. Overall, the respondents believed that the code generated by our method performed best across all three dimensions. Simultaneously, based on the actual needs of large-scale industrial development, they proposed three feasible directions for future optimization. Firstly, the current method only supports the generation of code for static pages and cannot recognize the polymorphism of elements or support the interaction logic between elements. Specifically, for UI elements with multiple states, Prototype2Code can only convert them into images, which still requires manual modification. Secondly, Prototype2Code supports only the generation of native HTML+CSS code. It is crucial to support for popular front-end development frameworks such as Vue and React, as well as cross-platform code generation. Lastly, the current method lacks interaction design. Developing a visual platform similar to CodeFun, which supports personalized operations by engineers, is a potential area for future research.

\section{Conclusion}

In this paper, we introduce Prototype2Code, an end-to-end approach designed for generating front-end code directly from UI design prototypes. By integrating prototype linting into the code generation workflow, Prototype2Code reduces the impact of structural flaws in design prototypes on the generation outcomes, thereby enhancing the robustness of the generation process. The rendered UI effects from Prototype2Code demonstrate superior visual quality, achieving an SSIM of 0.91, a PSNR of 21.15, and an MSE of 0.97e-2, significantly surpassing other existing methods. Our user study on code quality further provides an assessment from a development perspective. Overall, the code generated by Prototype2Code, due to the clear structure offered by the layout tree and the semantics obtained from type recognition, surpasses other methods in readability. Owing to the nested grouping and CSS generated by LLM, it also exhibits better availability, requiring only minimal manual adjustments to meet basic business requirements. Additionally, its compatibility with responsive design principles contribute to improved maintainability.

% Compared to existing commercial code generation platforms, Prototype2Code excels in addressing potential non-standard design issues within UI prototypes, incorporating fragmented elements detection and perceptual grouping into the code generation workflow, thereby boosting the method's robustness. To produce more readable code with clearer structure, Prototype2Code reconstructs the hierarchical information of design prototypes and employs Graph Neural Networks (GNN) for the intelligent recognition of UI elements. By supporting the flexbox layout model, it also accommodate responsive design requirements, ensuring that the generated front-end code is adaptable to devices of various sizes. For evaluation, we compared Prototype2Code with the commercial code platform CodeFun and the GPT-4v driven open-source tool Screenshot-to-code. In terms of the visual fidelity, Prototype2Code achieved the highest similarity with design prototypes with the fewest errors. Furthermore, in our user study involving five front-end engineers, participants unanimously concurred that Prototype2Code outperformed other methods in code readability, usability, and maintainability, requiring only minimal manual adjustments to fulfill basic business requirements. 

\bibliographystyle{asmeconf}  %% .bst file following ASME conference format. Do not change.
% \bibliography{asmeconf-sample}%% <=== change this to name of your bib file

%%%  APPENDICES  %%%%%%%%%%%%%%%%%%%%%%%%%%%%%%%%

%% Note that appendices will be "numbered" A, B, C, ... etc. Use \section, not \section*
%% Equations will be numbered sequentially following those in the paper. Do not reset the equation counter.

%% Here we use the optional argument to control the pdf bookmark and prevent errors.

%%%%%%%%%%%%%%%%%%%%%%%%%%%%%%%%%%%%%%%%%%%%%%%%%%%%%%%%%%%%%%%%%%%%%%%%%%%%%%%%%%%%%%%

\end{document}